\documentclass[10pt,twocolumn,aps,pre,superscriptaddress]{revtex4-1} %

\usepackage{graphicx}
\usepackage{dcolumn}
\usepackage{amsmath}
\usepackage{amssymb}
\usepackage{bm}
\usepackage{color}
\usepackage{mathrsfs}
\usepackage{multirow}
\usepackage{ulem}
\newcommand{\ud}{\mathrm{d}}

\begin{document}

\title{Lattice-Boltzmann Hydrodynamics of Anisotropic Active Matter}

\author{Joost de Graaf}
\email{jgraaf@icp.uni-stuttgart.de}
\affiliation{Institute for Computational Physics, University of Stuttgart, Allmandring 3, 70569 Stuttgart, Germany}

\author{Henri Menke}
\affiliation{Institute for Computational Physics, University of Stuttgart, Allmandring 3, 70569 Stuttgart, Germany}

\author{Arnold J.T.M. Mathijssen}
\affiliation{The Rudolf Peierls Centre for Theoretical Physics, 1 Keble Road, Oxford, OX1 3NP, United Kingdom}

\author{Marc Fabritius}
\affiliation{Institute for Computational Physics, University of Stuttgart, Allmandring 3, 70569 Stuttgart, Germany}

\author{Christian Holm}
\affiliation{Institute for Computational Physics, University of Stuttgart, Allmandring 3, 70569 Stuttgart, Germany}

\author{Tyler N. Shendruk}
\affiliation{The Rudolf Peierls Centre for Theoretical Physics, 1 Keble Road, Oxford, OX1 3NP, United Kingdom}

\date{\today}
\
\begin{abstract}
A plethora of active matter models exist that describe the behavior of self-propelled particles (or swimmers), both with and without hydrodynamics. However, there are few studies that consider shape-anisotropic swimmers and include hydrodynamic interactions. Here, we introduce a simple method to simulate self-propelled colloids interacting hydrodynamically in a viscous medium using the lattice-Boltzmann technique. Our model is based on raspberry-type viscous coupling and a force/counter-force formalism which ensures that the system is force free. We consider several anisotropic shapes and characterize their hydrodynamic multipolar flow field. We demonstrate that shape-anisotropy can lead to the presence of a strong quadrupole and octupole moments, in addition to the principle dipole moment. The ability to simulate and characterize these higher-order moments will prove crucial for understanding the behavior of model swimmers in confining geometries.
\end{abstract}

\maketitle

\section{\label{sec:intro}Introduction}

The field of active matter has seen a surge in interest from the scientific community during the last decade~\cite{ramaswamy10a,marchetti13a}. In particular, self-propelled particles (swimmers) have received a great deal of attention, due to recent breakthroughs that enable their fabrication on colloidal (1 nm - 1 $\mu$m) length scales~\cite{ismagilov02,paxton04a,ebbens10a,hong10a,sengupta12a,wang13a,sanchez15a} and their inherent out-of-equilibrium  nature~\cite{cates12a,Cates15a}. In such systems unexpected phenomena can occur,~\textit{e.g.}, giant number fluctuations, rectification, and collective motion~\cite{ramaswamy10a,hong10a,marchetti13a}. In addition, artificial swimmers have been connected to a wide range of practical applications, ranging from cancer therapy to soil remediation~\cite{lien99a,nelson10a,wang14a}, and countless naturally occurring self-propelled `particles' have been identified, including humans~\cite{Helbing00}, birds~\cite{Ballerini08}, fish~\cite{Katz11}, insects~\cite{Buhl06}, 
spermatozoa~\cite{Woolley03,Riedel05,Ma14}, bacteria~\cite{Sokolov07,Schwarz-Linek12,Reufer14}, and algae~\cite{Polin09,Geyer13}.

While the principle of autonomous motion is relevant across many orders of magnitude in size and speed, some of the most interesting behavior is found for colloidal swimmers suspended in a viscous medium. The low Reynolds number of these systems imposes that the motion of mechanical swimmers must be nonreciprocal, as put forward by Purcell's scallop theorem~\cite{purcell77}. Another class of swimmers are driven by phoretic mechanisms~\cite{ismagilov02,paxton04a,ebbens10a,hong10a,sengupta12a,wang13a,sanchez15a}. Catalytic reactions taking place heterogeneously over the particle's surface break the time-reversal symmetry for self-phoresing colloids. Both types of swimmer have force-free flow fields,~\textit{i.e.}, there is no monopole term (Stokeslet) to the flow field in the absence of external forces: only higher order hydrodynamic interactions (HI) are present in the system. This has important consequences for their interaction with each other~\cite{lauga09}, their environment~\cite{spagnolie12,zhu13,
uspal14,mathijssen16}, and tracer particles~\cite{pushkin13,Morozov14,mathijssen15}. 

A wide range of techniques is available to study the behavior of active particles using theory and simulations. The most common model is active Brownian model (ABM), which does not take HI into account. It has been applied to simulate shape-anisotropic particles, in bulk~\cite{wensink12}, under confinement~\cite{wensink08}, and in capturing geometries~\cite{kaiser13}. Hydrodynamics can be incorporated to first order, by introducing a point-like dipole interaction, either using lattice-Boltzmann (LB) simulations~\cite{nash08,nash10}, or using a Stokesian description~\cite{Hernandez-Ortiz05,saintillan07,swan11,singh15}. Universal aspects of swimmer dynamics are known to arise from these long-range hydrodynamics. Therefore, having models that can effectively simulate the correct long-range HI, without concern for specific short-range interactions is very valuable. 

Biological and artificial self-propelled particles often have a significant shape asymmetry,~\textit{e.g.}, sperm or bacteria~\cite{Woolley03,Riedel05,Ma14,Sokolov07,Schwarz-Linek12,Reufer14} and L-shaped colloids~\cite{kummel13}. This requires modification of the far-field (and near-field) hydrodynamic description that goes beyond the dipole term. That is, the shape anisotropy induces a series of higher-order multipole moments~\cite{spagnolie12}. The squirmer model~\cite{lighthill52,blake71} incorporates such higher-order HI. Squirmers have been studied using LB~\cite{pagonabarraga13,Li14,Lintuvuori15}, multi-particle-collision-dynamics (MPCD)~\cite{downton09,zoettl14,schaar15}, and other hydrodynamic solvers~\cite{Ishikawa08,Doostmohammadi12,zhu13,Molina13,Matas-Navarro14,Delmotte15}. The model has many appealing features, for instance, the ability to specify the exact hydrodynamic character of the swimmers and the possibility to add lubrication corrections~\cite{Ishikawa06}. Unfortunately, the squirmer 
model has only been extended to ellipsoidal swimmers~\cite{Delmotte15}, limiting its use to study the wide range of anisotropic shapes available.

Thus far, there have been very few studies of highly shape-anisotropic particles, where HI have been taken into account~\cite{Tao08,Elgeti10,lugli11,hu15,Wu15}. The models that have been considered are typically highly system specific, or too computationally expensive to simulate a large number of particles. An intermediate form is therefore required that takes into account HI and strikes a balance between the accurate simulation of shape-anisotropic particles and computational efficiency. The LB algorithm has been shown to efficiently simulate HI and is therefore considered ideally suited to this task. In particular, the simulations of Nash~\textit{et al.}~\cite{nash10} demonstrate that LB can readily simulate thousands of swimmers.

In this manuscript, we introduce a model with the aforementioned features. We simulate colloids of arbitrary shape by approximating them as clusters of spheres, see Fig.~\ref{fig:rasp}. These clusters are coupled to an LB fluid using the viscous coupling scheme introduced by Ahlrichs and D{\"u}nweg~\cite{ahlrichs99}, in which the friction depends on the relative velocity of particle and fluid. The effect of this coupling is the formation of a hydrodynamic hull around the points, which thus gain an effective hydrodynamic size~\cite{ahlrichs99}. Thereby, a solid particle can be modelled, which resembles a raspberry~\cite{lobaskin04} for a sufficient density of coupling points~\cite{fischer15,degraaf15b}. Self-propulsion is introduced by following the principles of Refs.~\cite{Hernandez-Ortiz05,saintillan07,nash08}: We assign a direction (unit) vector to the raspberry particle and apply a persistent force along this direction and an equal and opposite (counter) force to the fluid, see Fig.~\ref{fig:rasp}a. The 
location of the counter force determines the nature of the leading dipole moment and distinguishes pusher (extensile) from puller (contractile) swimmers. 

Using our model, we demonstrate that the anisotropy of the particle induces higher order multipole moments, in addition to the dipole moment that we impose. These multipole moments account for the flow of fluid around the object. We introduce a method to determine the magnitude of these multipole moments by means of a Legendre-Fourier analysis -- we limit ourselves to axisymmetric anisotropic swimmers here. This characterization technique is applicable beyond our raspberry swimmers, and may be of use in establishing the hydrodynamic nature of complex swimmers, for which the flow field has only been determined numerically~\cite{yoshinaga10,bickel13,degraaf15a,samin15}. We confirm that we obtain the correct leading-order dipole moment by considering the entrainment of a tracer particle in the flow field caused by the swimmer. This also serves as a proof of principle that our raspberry-swimmer model has applications in more complex settings than single-particle bulk simulations.

The raspberry swimmers introduced in this manuscript present a facile method, by which anisotropic swimmers can be modelled. It allows us to incorporate HI that go beyond the principle dipole and are essential to the accurate description of shape-anisotropic swimmers. The raspberry swimmers can be utilized to gain new insights into the behavior of anisotropic swimmers in various geometries, as we will further demonstrate in Ref.~\cite{degraaf16b}.

\section{\label{sec:methods}Methods}

In this section, we give an overview of the approaches followed to obtain the results presented in Section~\ref{sec:result}. We first outline the principles of the raspberry model and the construction of the raspberry swimmers. Next, we specify the Molecular Dynamics (MD) and LB parameters used in our work. We subsequently provide details on the simulations used to characterize the shape of the swimmers. This is followed by a discussion on the determination of the hydrodynamic moments via a Legendre-Fourier decomposition method, as well as by considering the entrainment of a tracer particle.

\subsection{\label{sub:raspmod}The Raspberry Model}

Ahlrichs and D{\"u}nweg~\cite{ahlrichs99} introduced a simple coupling scheme to simulate moving particles in an LB fluid without using Ladd (grid-based) bounce-back boundary conditions~\cite{ladd94}. In this approach, the particles are described as points that couple to the fluid through a frictional force, acting both on the solvent and on the solute, which depends on the relative velocity. A hydrodynamic hull forms around the points, which thus gain a finite hydrodynamic extent (effective hydrodynamic radius)~\cite{ahlrichs99}, due to this coupling and the interpolation of the force onto the LB grid. This method is particularly suited to simulate a monodisperse system of colloids where the far-field hydrodynamics dominate over the near-field contributions, which are typically not accurately captured. 

Lobaskin and D{\"u}nweg~\cite{lobaskin04} utilized the point coupling to simulate extended objects moving through a fluid by introducing the ``raspberry'' model. A larger particle -- compared to a single coupling point -- is modeled by discretizing the surface (and interior~\cite{fischer15}) of the particle. The method derives its name from this discretized nature of the surface, which resembles a raspberry, when represented by molecular-dynamics (MD) beads. When a sufficient number of points is used to couple to the fluid, the LB fluid inside the particle co-moves with the coupling points, thus modelling a hydrodynamically solid object. This raspberry coupling typically leads to an effective hydrodynamic extent of the object, that is larger than that of outmost coupling points~\cite{fischer15}. Despite the introduction of other methods of describing extended particles in a LB fluid, the raspberry method has remained popular, due to its simplicity.

\subsection{\label{sub:raspberry}Raspberry Swimmers}

\begin{figure*}
\begin{center}
\includegraphics[scale=1.0]{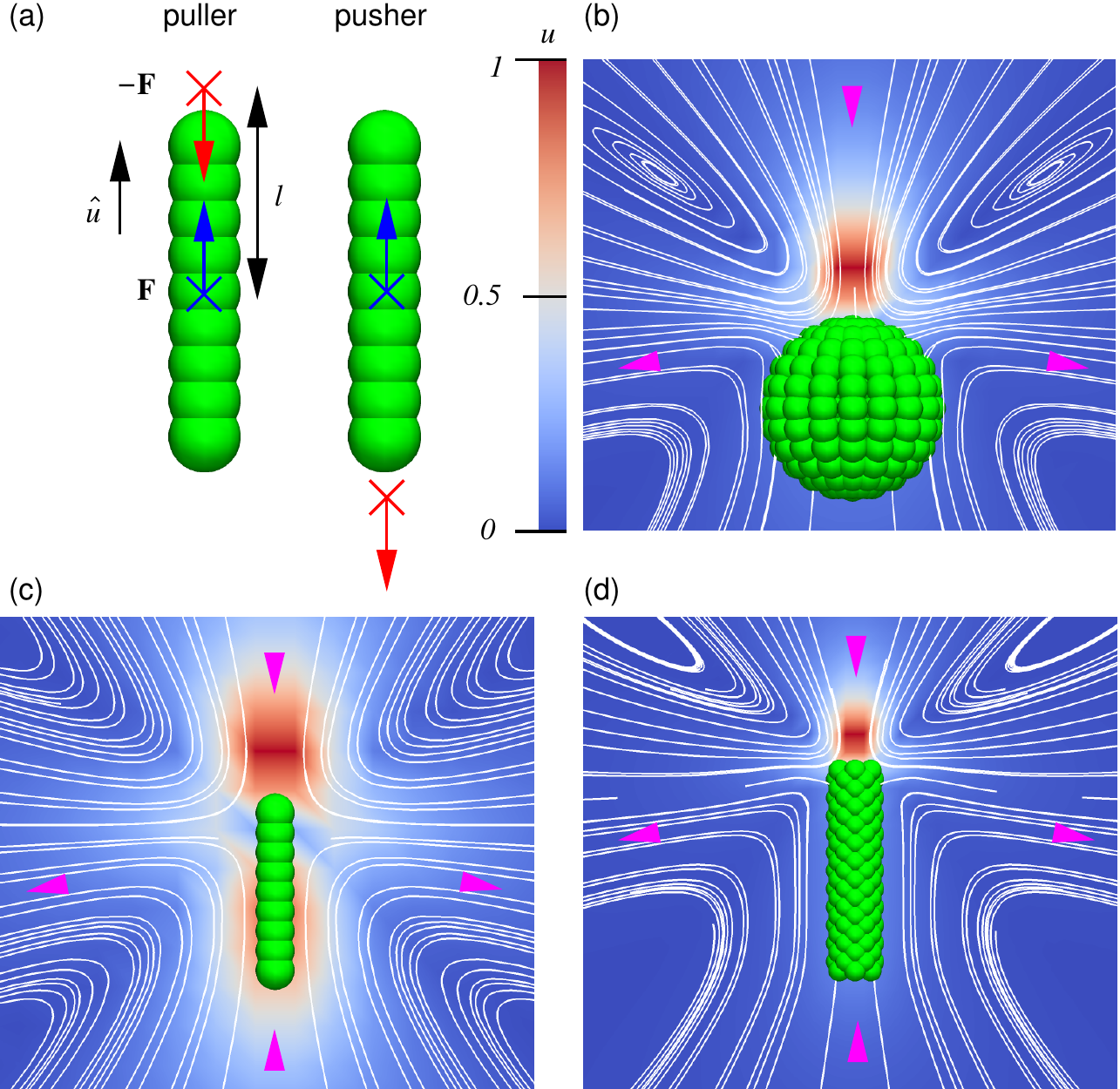}
\end{center}
\caption{\label{fig:rasp}(color online) The construction of raspberry swimmers. (a) A sketch of the construction of pusher and puller raspberry swimmers, in this case rods. The viscous coupling leads to an effective hydrodynamic radius, as indicated by the use of green spheres with a radius comparable to the effective one ($\sim0.5\sigma$). A force $\mathbf{F}$ (blue arrow) is applied to the central bead (blue cross) in the direction of the symmetry axis $\hat{u}$ (black arrow). A counter force $-\mathbf{F}$ (red arrow) is applied to the fluid at a point $l \hat{u}$ (red cross), with $l$ the dipole length. For $l>0$ the particle is a puller and for $l<0$ it is a pusher. (c-d) The flow field around puller raspberry swimmers. The normalized magnitude of the flow velocity in the lab frame is indicated by the coloring (red $\max \vert \mathbf{u}(\mathbf{r}) \vert = 1$, dark blue $\vert \mathbf{u}(\mathbf{r}) \vert = 0$) in a plane through the symmetry axis that is parallel to one of the box faces; only a part of 
the box is shown. The location of the counter-force point is clearly visible as a red region. The white curves are stream lines to the flow field and the magenta arrow heads indicate the direction of flow. We show three of our models: (b) the rod, (c) the sphere, and (d) the cylinder.}
\end{figure*}

In this manuscript, we study four different particle shapes using the so-called `raspberry' model for particle-fluid interactions~\cite{lobaskin04}. These are a point-particle, a sphere (Fig.~\ref{fig:rasp}b), a rod (Fig.~\ref{fig:rasp}a,c), and a cylinder (Fig.~\ref{fig:rasp}d). The details of the construction of generic raspberry particles are given in Ref.~\cite{fischer15}. We will only remark on several important construction differences and features of our models in the following. 

The point-particle is similar to the system of Nash~\textit{et al.}~\cite{nash08,nash10} and will serve as a reference to which we compare our anisotropic particles. It should be noted that the Ahlrichs and D{\"u}nweg coupling scheme~\cite{ahlrichs99} does not lead to rotational motion in a quiescent LB fluid (without an external torque), because there is no rotational coupling to the vorticity of the fluid flow. Nash~\textit{et al.} have suggested a simple model to introduce such rotations, which we find to be sensitive to lattice artifacts without external flow, even when a 3-point interpolation scheme is used. We therefore prefer to introduce rotational coupling by utilizing the properties of an extended raspberry model, which is known to accurately reproduce the desired rotational coupling~\cite{lobaskin04,fischer15,degraaf15b}.

We chose an axisymmetric distribution of $269$ coupling points for the sphere raspberry with two concentric shells to introduce internal coupling points. The shells contain $134$ points each, split up over $12$ semi-circles with $11$ equidistantly spread points and $2$ points at the pole; the central bead makes for $269$. This is different from the construction recipe provided in Ref.~\cite{fischer15}, where a random distribution of coupling points was used. We favored an axisymmetric distribution here, since the swimmer has a preferred direction, namely its direction of motion. An asymmetric (random) distribution leads to undesirable deviations from rectilinear motion in the absence of external torques. 

The rod and cylinder raspberry models simulate oblong particles. The rod is built up of 9 coupling points spaced $0.5\sigma$ apart over a line, with $\sigma$ the LB grid spacing ($\sigma$ is also the MD unit length). The cylinder consists of $161$ coupling points spread over $23$ groups of hexagonal disks (7 particles with distance $\sigma$), stacked alternatingly with a separation of $0.5\sigma$ along the axis. The rod is a simplified version of the cylinder, because the rod cannot experience fluid-flow induced rotation about its short axis in the standard coupling scheme for the same reasons that the point particle cannot rotate~\cite{ahlrichs99}. Moreover, the hydrodynamic coupling of the rod is substantially reduced with respect to that of the cylinder, such that the quality of the cylindrical surface that it approximates is limited, due to the low coupling-point density~\cite{fischer15}. Finally, it should be noted that the cylinder model has been constructed to be axisymmetric about the direction of 
motion.

All particles are made into swimmers by assigning a unit (direction) vector $\hat{u}$ to the particle, along its symmetry axis and originating in its center of mass (CM), see Fig.~\ref{fig:rasp}a. This $\hat{u}$ is updated according to the particle's orientation and its position (it co-moves). We apply a force $\mathbf{F}$ to the CM in the direction of $\hat{u}$ ($\mathbf{F} = F \hat{u}$) to cause the raspberry particle to move. We follow the approach of Nash~\textit{et al.} and apply a counter force $-\mathbf{F}$ to the fluid at a position $l\hat{u}$, with $l$ the dipole length that can be positive or negative, see Fig.~\ref{fig:rasp}a. Thereby, the swimmer reaches a terminal (swimming) velocity $U$ and its flow field is force free. For positive values of $l$ the swimmer is a puller and for negative values it is a pusher, see Fig.~\ref{fig:rasp}a. Our choice of the force coupling causes the dipole moment to be off center with respect to the CM. For the swimmer to obtain a reasonable $U$, we require $l > \sigma$.

\subsection{\label{sub:simulation}Simulation Parameters}

We used a graphics processing unit (GPU) based LB solver,~\cite{roehm12} that is attached to the MD software \textsf{ESPResSo}~\cite{limbach06a,arnold13a}. The GPU variant of LB implemented in \textsf{ESPResSo} utilizes a D3Q19 lattice and a fluctuating multi-relaxation time (MRT) collision operator~\cite{dhumieres02}. All of our simulations were performed in a quiescent (unthermalized) a LB fluid. We employed the LB viscous coupling of Ref.~\cite{ahlrichs99}, with a 3-point interpolation stencil~\cite{ladd94}. We found the 2-point interpolation to give rise to lattice artifacts,~\textit{e.g.}, oscillations in $U$ of over 20\% for the point particle.

Here, we employ the same LB parameters as in Refs.~\cite{fischer15,degraaf15b}, since these lead to faithful reproduction of the Stokesian mobility tensor -- both in bulk and under confinement. We set the fluid density to $\rho = 1.0 m_{0}\sigma^{-3}$, the lattice spacing to $1.0\sigma$, the time step to $\Delta t = 0.005 \tau$ ($\tau$ is the MD time unit), the (kinematic) viscosity to $\nu = 1.0 \sigma^{2}\tau^{-1}$, and the bare particle-fluid friction to $\zeta_{0} = 25 m_{0}\tau^{-1}$, with $m_{0}$ the MD mass unit. We refer the reader to Ref.~\cite{fischer15} for a detailed description of the dimensionless numbers that specify the fluid properties to which these choices correspond.

On the MD level, the raspberry particles are allowed to freely move and rotate, unless otherwise specified. All the forces acting on the MD beads are transferred to the central bead \textit{via} the virtual sites (rigid bonds). To stabilize the simulation for the bare friction coefficients used, we set the (bare) mass and rotational inertia of the raspberry; these quantities should not be confused with the virtual mass of the body in a fluid, see, \textit{e.g.}, Ref.~\cite{Zwanzig75} for the definition. The mass and rotational inertia tensor are based on the particle's dimensions and the fluid mass density, and must be chosen reasonably to ensure the stability of the algorithm. The values of the imposed physical quantities are listed in Table~\ref{tab:params}.

\begin{table}
\begin{ruledtabular}
\begin{tabular}{c|c|c|c|c|c||c|c}
shape     & $R/\sigma$ & $H/\sigma$ & $M/m_{0}$ & $I_{\parallel}$ & $I_{\perp}$ & $R_{h}/\sigma$ & $H_{h}/\sigma$ \\
\hline
point     &    -       &          - &       4.8 &             4.8 &         4.8 &           0.56 &       - \\     
sphere    &  2.5       &          - &       66. &             160 &         160 &           3.1  &       - \\
rod       &    -       &        2.0 &       7.1 &             1.5 &          17 &           0.87 &     2.9 \\
cylinder  &  1.0       &        5.5 &       85. &             95. &        1100 &           1.6  &     6.1 \\
\end{tabular}
\end{ruledtabular}
\caption{\label{tab:params}List of the properties of our raspberry swimmers. From left to right, the particle shape, the imposed radius $R$ and half length $H$, the mass of the particle $M$, the moment of inertia parallel to the axis of symmetry $I_{\parallel}$ (with unit $m_{0}\sigma^{2}$), the moment perpendicular to this axis $I_{\perp}$ (with unit $m_{0}\sigma^{2}$), and the measured effective hydrodynamic radius $R_{h}$ and half length $H_{h}$. We used the MD units $\sigma$ (unit length) and $m_{0}$ (unit mass) to de-dimensionalize our parameters.}
\end{table}

\subsection{\label{sub:shape}Raspberry Characterization}

We employed the methods of Refs.~\cite{fischer15,degraaf15b} to characterize the hydrodynamic properties of the raspberry particles. For the point and sphere swimmer, the methods of Refs.~\cite{fischer15,degraaf15b} could be directly applied to determine the effective hydrodynamic radius $R_{h}$. For the rod and cylinder swimmer, we used the formalism developed in Ref.~\cite{fischer15} for a dumbbell. The theoretical expressions for the hydrodynamic mobility tensor (HMT) of a cylinder segment~\cite{tirado84} were used to extract the effective hydrodynamic radius $R_{h}$ and half length $H_{h}$ of the rod and cylinder.

There are two fit parameters for a rod-shaped particle ($R_{h}$ and $H_{h}$) that can be extracted from the measurements, to which the extrapolated bulk HMT must be fit. In order to determine these two parameters simultaneously, we minimized the following functional
\begin{align}
\nonumber     f(H,R) &= \big( ( \mu_{\parallel}^{m} - \mu_{\parallel}(H,R) )^{2} + ( \mu_{\perp}^{m} - \mu_{\perp}(H,R) )^{2} \\
\label{eq:min}       & + ( \mu_{r}^{m} - \mu_{r}(H,R) )^{2} \big),
\end{align}
where the superscript `$m$' signifies the measured quantity, $\mu_{\parallel}$ is the translational mobility parallel to the symmetry axis, $\mu_{\perp}$ the translational mobility perpendicular to this axis, and $\mu_{r}$ the mobility associated with reorientation of the symmetry axis. To minimize internal inconsistency we use a single fit parameter for both $H_{h} = H + \Delta$ and $R_{h} = R + \Delta$. That is, we minimize $f(H + \Delta,R + \Delta)$, with $H$ and $R$ the imposed half-length and radius. This also ensures a well-definedness of the result for $R = 0$ (the rod), since we have $\Delta > 0$ which eliminates the divergences in the logarithmic terms of Ref.~\cite{tirado84}. The value of $\Delta$ found for the rod is then its effective hydrodynamic radius and should be comparable to the effective radius of a point. This is indeed the case, as can be inferred from the measured parameters listed in Table~\ref{tab:params}.

\subsection{\label{sub:moments}Moment Characterization}

The hydrodynamic properties of the raspberry swimmers are assessed by placing a single swimmer in the center of a cubic box with side length $L = 150 \sigma$ and periodic boundary conditions, that is centered on the origin and axis aligned. This length $L$ is a trade-off between simulation speed and minimization of periodicity effects. We let the swimmer move along the $z$-axis in the positive direction and allow a steady-state flow field to set in and take a snapshot, see Fig.~\ref{fig:rasp}b-d. We determine the distance $d$ travelled by the swimmer and we shift the measured flow field by this distance, such that the swimmer's CM is located in the origin. 

The flow fields display a clear hydrodynamic dipole. The near-field curvature of the flow-lines is due to the finite separation of the force and counter-force point. Note that the flow lines mostly pass around the extended raspberries (sphere and cylinder), indicating that the viscous coupling indeed causes a hydrodynamic hull to form around the raspberry. Those that pass through the object are mostly due to the result being shown in the laboratory frame, although the extended raspberries can have a slight porosity~\cite{fischer15}. It is the flow around the object, coupled to the shape-anisotropy, that is responsible for the presence of higher-order moments in the swimmer's flow field. Finally, far away from the swimmer (not visible on the scale of Fig.~\ref{fig:rasp}b-d) the flow lines are closed due to the periodic boundary conditions.

The flow field is Legendre-Fourier (LF) decomposed into modes to determine these hydrodynamic moments. First, we transform the fluid velocity field $\mathbf{u}(\mathbf{r})$ from Cartesian coordinates to spherical polar coordinates, with $\phi$ the azimuthal and $\theta$ the polar angle. The azimuthal $\hat{\phi}$ component is zero due to symmetry and can be ignored, leaving the radial $\hat{r}$ and polar (tangential) $\hat{\theta}$ component. This flow field is then cylindrically averaged around the $z$-axis to arrive at expressions of the form $u_{r}(r,\theta)$ and $u_{\theta}(r,\theta)$ for the radial and tangential components of the flow field, respectively. Finally, we project out the $\theta$ dependence using LF decomposition
\begin{align}
\label{eq:LFr} u_{x,n}(r) & = \frac{2n + 1}{2} \int_{0}^{\pi} \sin(\theta) L_{n}(\cos(\theta)) u_{x}(r,\theta) \ud \theta ,
\end{align}
with $n$ the index of the mode, $L_{n}$ the Legendre polynomial of order $n$, and $x = r$ or $x = \theta$ depending on the component. For the series of mode functions $u_{r,n}(r)$ and $u_{\theta,n}(r)$, we fit the long-range decay using a power-law, see,~\textit{e.g.}, Fig.~\ref{fig:LF}.

We compare the power-law decay obtained in our modes to the expected decay of the monopole (st, Stokeslet), dipole (di, Stresslet), source-dipole (sd), quadrupole (qu), octupole (oc), and source-octupole (so) moments. We found that there are no monopole and source terms~\footnote{The absence of source terms is specific to the LB force/counter-force raspberry swimmer. That is, there are no sources or sinks of the LB fluid. However, other models such as the squirmer~\cite{lighthill52,blake71} can have source-dipole contributions that must be considered in fitting the data.} ($\mathbf{u}_{\mathrm{st}}$, $\mathbf{u}_{\mathrm{sd}}$, and $\mathbf{u}_{\mathrm{so}}=\mathbf{0}$). Thus the expressions for the flow fields generated by a swimmer that is centered at the origin and pointing along the $\hat{z}$ direction are given by a sum of
\begin{align}
\label{eq:dipole} \mathbf{u}_{\mathrm{di}}(r,\theta) &= \frac{A}{r^{2}} \left( \frac{1 + 3\cos(2\theta)}{2} , 0 \right);\\
\nonumber         \mathbf{u}_{\mathrm{qu}}(r,\theta) &= \frac{B}{r^{3}} \bigg( -\frac{\cos(\theta) + 3\cos(3\theta)}{2} , \\
\label{eq:quadru}                                  &\phantom{= \frac{B}{r^{3}} \big( -} \frac{\sin(\theta) - 3 \sin(3\theta)}{8} \bigg);\\
\nonumber         \mathbf{u}_{\mathrm{oc}}(r,\theta) &= \frac{C}{r^{4}} \bigg( \frac{3 + 4\cos(2\theta)+25\cos(4\theta)}{16} , \\
\label{eq:octapo}                                  &\phantom{= \frac{C}{r^{4}} \bigg( -} \frac{-2\sin(2\theta)+5\sin(4\theta)}{8} \bigg),
\end{align}
up to fourth order. Here, the coefficients $A$, $B$, and $C$ give the strength and we list the components $\hat{r}$ (first entry) and $\hat{\theta}$ (second entry). When the first four hydrodynamic moments are LF decomposed, we obtain a series of power-law decays.

\begin{table}
\begin{ruledtabular}
\begin{tabular}{c|c||c|c|c|c|c|c}
\multicolumn{2}{c||}{name}                      &           st &        di &            sd &              qu &              oc &           so \\
\hline
\multicolumn{2}{c||}{decay}                     &     $r^{-1}$ &  $r^{-2}$ &      $r^{-3}$ &        $r^{-3}$ &        $r^{-4}$ &     $r^{-4}$ \\
\hline
n                              &          comp  &         pref &      pref &          pref &            pref &            pref &         pref \\
\hline
\hline
\multirow{2}{*}{0}             &      $\hat{r}$ &            0 &         0 &             0 &               0 &              0 &             0 \\
                               & $\hat{\theta}$ & $-\pi/2^{2}$ &         0 &   $\pi/2^{2}$ &     $\pi/2^{5}$ &              0 &             0 \\
\hline
\multirow{2}{*}{1}             &      $\hat{r}$ &            2 &         0 &             2 &           $2/5$ &              0 &             0 \\
                               & $\hat{\theta}$ &            0 &         0 &             0 &               0 &  $-3\pi/2^{5}$ & $-3\pi/2^{3}$ \\
\hline
\multirow{2}{*}{2}             &      $\hat{r}$ &            0 &         2 &             0 &               0 &         $-6/7$ &          $-2$ \\
                               & $\hat{\theta}$ & $5\pi/2^{5}$ &         0 & $-5\pi/2^{5}$ &  $-25\pi/2^{7}$ &              0 &             0 \\
\hline
\multirow{2}{*}{3}             &      $\hat{r}$ &            0 &         0 &             0 &         $-12/5$ &              0 &             0 \\
                               & $\hat{\theta}$ &            0 &         0 &             0 &               0 & $203\pi/2^{9}$ &  $7\pi/2^{5}$ \\
\hline
\multirow{2}{*}{4}             &      $\hat{r}$ &            0 &         0 &             0 &               0 &         $20/7$ &             0 \\
                               & $\hat{\theta}$ & $9\pi/2^{8}$ &         0 & $-9\pi/2^{8}$ & $387\pi/2^{12}$ &              0 &             0 \\
\end{tabular}
\end{ruledtabular}
\caption{\label{tab:modes}The Legendre-Fourier (LF) mode decomposition of the first hydrodynamic moments for a swimmer that can have an arbitrarily complex flow field. The first row lists the moments: monopole (st, Stokeslet), dipole (di, Stresslet), source-dipole (sd), quadrupole (qu), octupole (oc), and source-octupole (so). The second row provides the decay. The third row labels the content of the rest of the table, from left to right: the mode, the radial/tangential component, and the prefactors. We only specify modes up to $n=4$ here. The table is used to determine a specific LF mode by combining the prefactor with the decay and relevant strength coefficient. For example, the 2nd mode of the tangential component of the quadrupole is given by $u_{\theta,2}^{\mathrm{qu}}(r) = -25 \pi B/(2^{7}r^{3})$.}
\end{table}

Table~\ref{tab:modes} shows the mode decomposition and the following observations can be made. (i) Every moment has a `unique' power-law decay: st $r^{-1}$, di $r^{-2}$, qu $r^{-3}$, oc $r^{-4}$, \textit{etc.}; provided that the source terms can be ignored. (ii) It is sufficient to consider $u_{\theta,0}$ and check for $r^{-1}$ decay to determine if the system is force free. (iii) The dipole is the only moment that has exactly one nonzero entry in its LF spectrum. Thus $A$ can be extracted by fitting $r^{-2}$ to the $u_{r,2}$ mode. (iv) For the quadrupole moment, the $u_{r,3}$ mode is ideally suited to determine the factor $B$, as there is no $n=3$ component to the source-dipole. (v) If we subtract the measured quadrupole moment from the $u_{\theta,2}$ mode then we find no remaining $r^{-3}$ decay, proving the absence of the source dipole moment. (vi) The strength $B$ of the quadruple moment can be double checked by using any of the other nonzero entries, provided that the source-dipole term can be excluded. (v) 
Finally, a similar approach can be followed to establish the value of $C$ for the octupole moment (by fitting $r^{-4}$ to $u_{r,4}$) and the higher-order moments.

It should be noted that there are short-ranged monopolar signatures, because point forces are used in the raspberry force/counter-force scheme. However, there is no long-range ($r \gg \max(H,R)$) decay that is proportional to $r^{-1}$. In fact, all the decays must be measured sufficiently far away from the swimmer, as the 3-point interpolation and finite size of the object will significantly modify the near-field shape of the decay, as we will see. Another source of error, when fitting, is the periodicity of the LB simulation domain. Care must be taken to only use the decay sufficiently far from the edge of the periodic simulation box. In practice, this limits the regime over which the decay can be fitted but these limitations are apparent from the data.

Finally, it is possible to use the LF mode decomposition to extract moments in other swimmer bases. We have used this method to extract a series of coefficients for the squirmer model~\cite{lighthill52,blake71}. The result is a set of coefficients that is analogous to the above moment decomposition and may be used to map our models onto LB simulations of squirmers -- we do not provide these coefficients here. Interestingly, the near-field correspondence between the raspberry flow field and that of the matched squirmer is not substantially improved, even for a spherical particle. We attribute this to the difference in method of achieving self-propulsion. That is, squirmers have a predefined surface slip velocity, which inadequately captures the presence of the counter-force point in the near field.

\subsection{\label{sub:entrain}Entrainment Matching}

We confirmed that our LF decomposition gives reasonable values for the strength of the dipole moment, by considering the entrainment of a tracer particle in the flow field of the swimmer. Again, we performed our simulations in a cubic box with edge length $L=150\sigma$, with periodic boundary conditions. The swimmer is initially positioned at $(0,0,-L/2)$ pointing in the $\hat{z}$ direction, and the tracer is placed at $(15\sigma,0,0)$. The minimum swimmer-tracer separation of $15\sigma$ gives reasonable results that are not too strongly affected by periodicity. As the swimmer moves from one edge of the box to the other, the tracer is advected (entrained) by the swimmers flow field. A sketch of this situation is provided in Fig.~\ref{fig:entr}.

The expression for the dipole moment can be used to numerically solve for the trajectory of a tracer using the form
\begin{align}
\label{eq:trace} \frac{ \partial \mathbf{r}_{\mathrm{tr}} }{ \partial t } &= \mathbf{u}_{\mathrm{di}}(\mathbf{r}_{\mathrm{tr}} - \mathbf{r}_{\mathrm{sw}}),
\end{align}
where $\mathbf{r}_{\mathrm{tr}}$ is the position of the tracer, $\mathbf{r}_{\mathrm{sw}}$ the position of the swimmer, and $\mathbf{u}_{\mathrm{di}}$ the dipole flow field generated by the tracer. Since the trajectory of the swimmer is a straight line and it moves at a constant speed, Eq.~\eqref{eq:trace} becomes a differential equation in terms of the tracer position only, which can be numerically evaluated. The theoretically predicted trajectory can then be fitted to the trajectory observed in the LB simulations for strength of the dipole moment, which we denote by $A^{\ast}$ to differentiate it from the value obtained from the mode decomposition.

The trajectory of the tracer has a characteristic concave triangular shape~\cite{pushkin13}; an example is shown in Fig.~\ref{fig:entr} for a puller. This shape is formed as follows (in the case of a puller). First, the tracer is pulled towards the swimmer, as the tracer is in front of the swimmer (i in Fig.~\ref{fig:entr}). Then the tracer is pushed away from the swimmer, when it is alongside of the swimmer (ii). Finally, when the swimmer moves away from the tracer, the tracer is again pulled towards the swimmer (iii). Here, we have accounted for the effects of periodic boundary conditions on the entrainment trajectories, which are minor.

\section{\label{sec:result}Results}

In this section we present our results. We first discuss the reasons for choosing specific values for the force/counter-force point separation $l$. This is followed by an analysis of our LF-mode decomposition applied to the raspberry particles that we constructed. Finally, we provide results for the entrainment experiment that we performed to verify our mode decomposition.

\subsection{\label{sub:mobility}Swimming Speed and the Counter-Force Point}

We first measured the hydrodynamic mobility of our particles to determine their shape and size using the procedures outlined in Section~\ref{sub:shape}, see Table~\ref{tab:params}. Please refer to Refs.~\cite{lobaskin04,fischer15} for additional details on the way in which the effective size and shape of the raspberry particles can be established. Using our fitted shape parameters, the mass and rotational inertia can be determined using the standard expressions for cylindrical and spherical objects, assuming a mass density that is the same of that of the fluid, see Table~\ref{tab:params}. This ensures the stability of the simulation~\cite{fischer15}. To verify that our anisotropic distribution of coupling points did not give the sphere an anisotropic hydrodynamic mobility tensor, we measured the sphere's mobility around different axes. We found the mobility to be isotropic to within an acceptable tolerance: all measurements gave mobilities that differed by at most 2\%.

\begin{table}
\begin{ruledtabular}
\begin{tabular}{c|c|c|c|c|c|c|c|c}
shape                      & $l/\sigma$ &                      F & $U$       &      $U_{\mathrm{pr}}$ &       $A$ & $A^{\ast}$ &       $B$ &     $C$ \\
$\times$                   &          1 &                      1 & $10^{-3}$ &              $10^{-3}$ & $10^{-2}$ &  $10^{-2}$ & $10^{-2}$ &       1 \\
\hline
\multirow{2}{*}{point}     &      $1.0$ &  \multirow{2}{*}{0.37} &     $20.$ & \multirow{2}{*}{$35.$} &    $-1.4$ &     $-1.4$ &       N/A &     N/A \\
                           &     $-1.0$ &                        &     $20.$ &                        &     $1.4$ &      $1.4$ &       N/A &     N/A \\
\hline
\multirow{2}{*}{sphere}    &      $3.5$ &   \multirow{2}{*}{3.8} &     $3.5$ & \multirow{2}{*}{$67.$} &    $-10.$ &     $-10.$ &       N/A &  $-3.1$ \\
                           &     $-3.5$ &                        &     $3.3$ &                        &     $10.$ &      $10.$ &       N/A &   $3.1$ \\
\hline

\multirow{2}{*}{rod}       &      $2.7$ &  \multirow{2}{*}{0.21} &     $2.5$ & \multirow{2}{*}{$3.4$} &    $-1.3$ &     $-1.4$ &     $3.7$ & $-0.11$ \\
                           &     $-2.7$ &                        &     $2.5$ &                        &     $1.3$ &      $1.4$ &     $3.7$ &  $0.11$ \\
\hline

\multirow{2}{*}{cylinder}  &      $6.5$ &  \multirow{2}{*}{0.60} &    $0.99$ & \multirow{2}{*}{$4.8$} &    $-2.7$ &     $-3.2$ &     $23.$ &  $-2.1$ \\
                           &     $-6.5$ &                        &     $1.0$ &                        &     $2.7$ &      $3.2$ &     $23.$ &   $2.1$ \\
\end{tabular}
\end{ruledtabular}
\caption{\label{tab:active}The imposed and measured properties of our LB raspberry swimmers. The table provides the shape, the position $l$ at which the counter force is applied (positive for a puller and negative for a pusher), the force $F$ in units of $m_{0} \sigma/\tau^{2}$, the measured velocity $U$ of the swimmer in units of ($\sigma/\tau$), the velocity $U_{\mathrm{pr}}$ in units of ($\sigma/\tau$) that is predicted on the basis of force and mobility, the dipole strength $A$ from the LF decomposition in units of ($\sigma^{3}/ \tau$), the dipole strength $A^{\ast}$ as measured in the entrainment experiment, and $B$ and $C$ the quadrupole and octupole strength from the LF decomposition in units of ($\sigma^{4}/ \tau$ and $\sigma^{5}/\tau$), respectively. The entries N/A indicate that a certain moment was not found.}
\end{table}

We varied the dipole lengths and force values to study their impact on the velocity and dipole strength. Provided the counter-force point is sufficiently far from the swimmer, one would expect the velocity of the swimmer to be dominated by the force applied to it directly and its hydrodynamic mobility, with the counter-force point having almost no effect. However, when the counter-force point is close to the swimmer, it starts to influence the measured velocity. In all cases, the velocity decreases with respect to that of a swimmer where the counter force is applied far away. When the counter-force point is too close, the effective swimming speed is negligible. This is also partially due to the 3-point interpolation applying the counter-force directly inside the volume occupied by the raspberry particle. Fortunately, when the counter-force is applied more than 1 grid spacing away from the closest coupling point, we found that the velocity $U$ of the swimmer is dominated by the mobility $\mu_{\parallel}$ 
and the value of $F$, i.e., $U = \mu_{\parallel} F$, and reasonable swimming speeds can be obtained. Table~\ref{tab:active} provides the swimming parameters (specifically the value of $l$) that we used for the remainder of our investigation. Our choice is based on a trade-off between stability and speed of the swimmer (or equivalently, overall run time of the simulation).

The values of the measured and predicted speed ($U_{\mathrm{pr}} = \mu_{\parallel} F$) in Table~\ref{tab:active} illustrate the speed reduction due to the counter-force being applied relatively close to the swimmer. For a sphere the effect is most pronounced, since we have $U_{\mathrm{pr}} = F/(6\pi \eta R_{h}) \approx 6.7\;10^{-2}\;\sigma/\tau$, where $R_{h}$ is the effective hydrodynamic radius (as discussed in Section~\ref{sub:raspmod}), and we observe $U = 3.5\;10^{-3}\;\sigma/\tau$. For the point, rod, and cylinder, on the other hand, $U_{\mathrm{pr}}$ is comparable to $U$, as these shapes experience less of an effect of the counter force. Also note that there can be a measurable difference between the speed of a pusher and a puller for the same raspberry particle, see the sphere entry in Table~\ref{tab:active}. This difference is caused by the asymmetry in the way the forces are applied for the two types of swimming, as shown in Fig.~\ref{fig:rasp}a, and is most strongly revealed for the simulations 
where the counter-force point is close to the swimmer. Finally, it should be pointed out that we averaged $U$ over several periods of the oscillations resulting from the lattice interpolation, see Ref.~\cite{fischer15} for a discussion. The deviation from this average was found to be limited to less than 5\%.

\subsection{\label{sub:LFres}The Legendre-Fourier Decomposition}

\begin{figure}
\begin{center}
\includegraphics[scale=1.0]{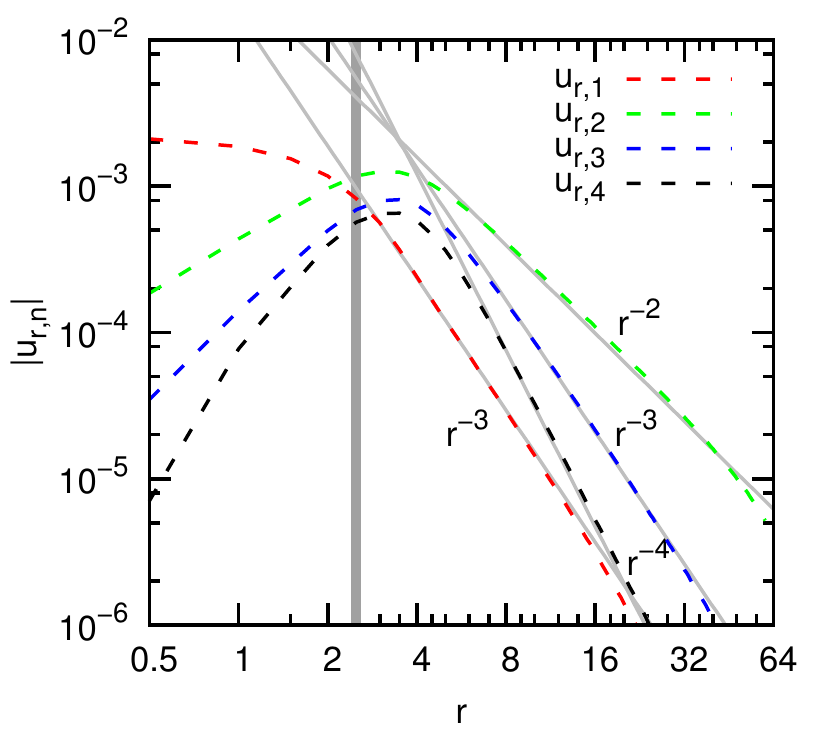}
\end{center}
\caption{\label{fig:LF}(color online) Legendre-Fourier mode decomposition of the flow field around a puller-type rod-shaped raspberry swimmer. Log-log plot of the first four radial modes obtained by our averaging and decomposition procedure: $u_{r,1}$ (red), $u_{r,2}$ (green), $u_{r,3}$ (blue), and $u_{r,4}$ (black). The dashed curves show the LB data, while the gray curves show the power-law fits to this data. The thick gray vertical line indicates $r = l$.}
\end{figure}

Figure~\ref{fig:LF} shows a representative example of the LF decomposition of the flow fields for the specific case of the rod-shaped puller. There are several points of interest. (i) The long-range decay of each mode follows a power law. These decays are fitted using $r^{-2}$, $r^{-3}$, and $r^{-4}$, respectively, showing that the flow field can be well approximated by a dipole, quadrupole, and octupole moment, to fourth order. (ii) For $u_{r,1}$ Table~\ref{tab:modes} predicts a sum of $r^{-n}$ terms, with $n=1,3$. However, there is clearly no $r^{-1}$ term, signifying that there is no long-range monopole moment, and the remaining decay is well captured using the $r^{-3}$ term only. (iii) Deviations far from the particle can be attributed to the periodicity of the simulation domain, but are limited, as can be seen. (iv) The small $r$ deviations from the expected power laws mark the onset of the near-field region in the immediate vicinity of the swimmers. For $r < l$, as indicated using the thick dark-gray 
line in Fig.~\ref{fig:LF}, the projection onto LF modes integrates over parts of the fluid where the particle is present. For $r \approx l$ the maximum is achieved, due to the counter-force point being included in the projection. (v) As explained in Section~\ref{sub:moments} the $u_{r,1}$ and $u_{r,3}$ curves can be fitted to determine the value of $B$ independently. (vi) The value of $C$ was extracted from $u_{r,4}$ and verified against other modes that decay as $r^{-4}$. Table~\ref{tab:active} lists the moments that were obtained using the decomposition. The error on the fit is very small, but there are numerous sources of systematic error (interpolation, averaging, etc.). From our analysis, we consider an estimated 15\% error to be justified, which we arrive at by considering the value obtained for the various modes.

We found quadrupole moments for both the rod and the cylinder using our LF decomposition. For the point swimmer, there was no sign of a quadrupole moment, since the viscous coupling and application of the counter force lead to a symmetric (albeit off-center from the bead) force configuration. Surprisingly, for the sphere we could not establish a quadrupole moment. The extent of the sphere, coupled with the off-center force/counter-force scheme that we used, should result in a quadrupole moment. However, it is likely that it was too small to be measured, despite the application of high forces for this swimmer. This leads us to conclude that the shape anisotropy of the rod and cylinder is the primary reason behind the large quadrupole moment. Spagnolie and Lauga~\cite{spagnolie12} relate the appearance of the quadrupole moment to a length asymmetry between the body and flagellum in a mechanically propelled swimmer. This interpretation lends itself to our data, as the shape anisotropic particle applies force to 
the fluid over a much greater length than the counter-force point does. However, Spagnolie and Lauga also claim that the size of the body induces a source dipole, which we do not observe in our results. Finally, we found an octupole moment for all particles, save the point swimmer (as expected). This implies that our model swimmers can be described by a series of force multipoles only, rather than a combination of force and source multipoles.

\subsection{\label{sub:entrres}Tracer Entrainment by a Raspberry Swimmer}

\begin{figure}
\begin{center}
\includegraphics[scale=1.0]{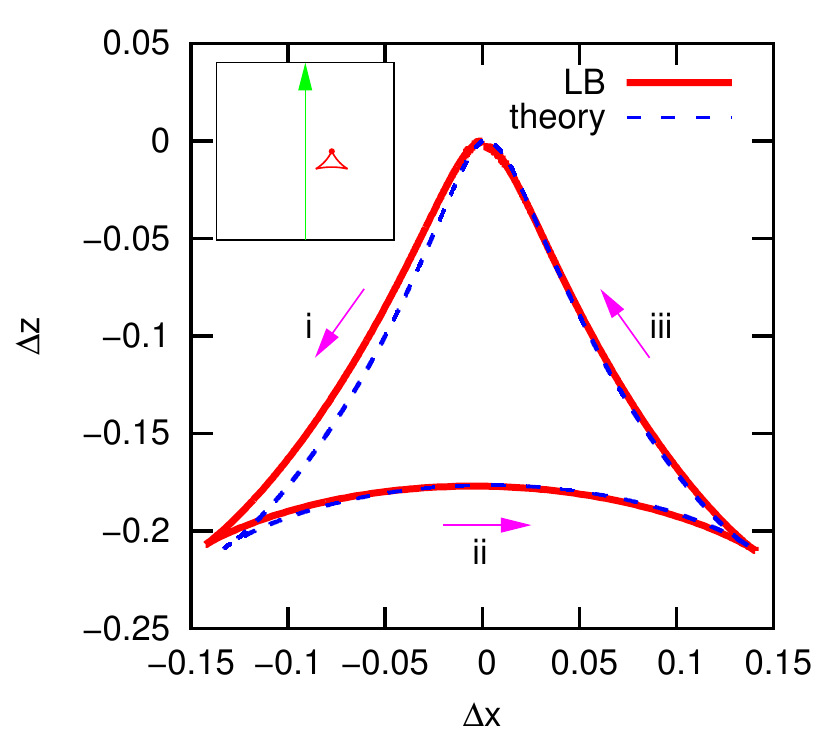}
\end{center}
\caption{\label{fig:entr}(color online) Entrainment curve for a tracer in the flow field of a rod-shaped raspberry swimmer. The parameters $\Delta x$ and $\Delta z$ show the amount of deviation from the tracer's original position as the swimmer, in this case a puller, moves through the box. The red curve shows the result of our LB simulations, while the blue dashed curve shows the fitted theoretical result. The numbers and magenta arrows indicate the way in which this curve is traversed by the tracer. The inset shows a sketch of the trajectory of a puller-type swimmer (green arrow) and the entrainment of the tracer (red dot) that this effects.}
\end{figure}

To verify our mode decomposition and demonstrate the utility of the raspberry-swimmer model, we considered the entrainment of a tracer particle in the fluid flow field generated by a swimmer, as explained in Section~\ref{sub:entrain}. Figure~\ref{fig:entr} shows a representative example of such an entrainment curve, in this case for the rod-shaped puller with a fitted coefficient $A^{\ast}$. The value of $A^{\ast}$ could be fitted with an estimated error of 15\%, due to the asymmetry in the curve. The curve is far less sensitive to the value of the quadrupole and octupole moment, because of their faster decay, which made it difficult to establish these moments from the curve with any accuracy. A much smaller tracer-swimmer separation would be required to measure the quadrupole moment and an even smaller separation for the octupole moment. However, this too introduces difficulties due to the near-field deviations from the power-law decay, as shown in Fig.~\ref{fig:LF}. 

Retardation effects cause the deviation observed on the left-hand side of Fig.~\ref{fig:entr}; part (i) of the trajectory. That is, the part of the trajectory for which the swimmer moves towards the particle. However, as the swimmer starts off in a quiescent fluid, it takes a finite simulation time for the steady-state flow field to be established, which is reflected in that part of the trajectory. Nevertheless, the fact that we can reproduce far-field tracer trajectories with a high level of accuracy indicates that our method can be successfully applied to more complex situations. 

\section{\label{sec:conc}Conclusion and Outlook}

Summarizing, we have introduced a new model to simulate anisotropic self-propelled colloids with hydrodynamic interactions utilizing the lattice Boltzmann method. Our LB model is based on the raspberry-type viscous coupling method introduced in~\cite{ahlrichs99,lobaskin04} and recently re-examined in detail in~\cite{fischer15,degraaf15b}. The raspberry particles are made to move by applying a force along a unit vector that describes their orientation. The correct force-free flow field is achieved by applying an opposing counter force to the fluid, see Fig.~\ref{fig:rasp}a. This force/counter-force formalism is similar to the ones introduced in Refs.~\cite{Hernandez-Ortiz05,saintillan07,nash08}. However, we go beyond the level of description presented there to introduce the particle shape and size. 

We verified that our raspberry swimmers model self-propelled colloids, by considering four basic shapes: a point, a sphere, a rod, and a cylinder, see Fig.~\ref{fig:rasp}. We discussed the creation of these swimmers, as well as the limitations of our method in detail. We introduce and carefully detail a Legendre-Fourier mode decomposition of the steady-state flow field (in bulk). This LF decomposition allowed us to determine the hydrodynamic moments of our swimmers, by fitting the mode-space decays with characteristic power-laws. The exponent of the decay for a specific mode is characteristic of a certain hydrodynamic moment. Using this formalism, we found that there is no monopole moment, as physically required for a force-free swimmer. The strengths for the series of higher-order moments were determined up to the octupole term. Our LF decomposition formalism is sufficiently generic to be applied to other swimmer bases as well and, for example, can be used to obtain the coefficient list for a squirmer. 

To validate our LF decomposition result, as well as have proof-of-concept application of our simulation method, we considered the entrainment of a tracer particle in the flow field of a passing swimmer. The entrainment curve allowed us to verify the dipole moment predicted by the decomposition. The quadrupole and octupole moments could not be verified in this fashion, due to their faster decay as well as near-field discretization artifacts. Fortunately, using the LF modes allows for internal verification of their strength. We observed that anisotropy introduces a strong quadrupole moment into the flow field surrounding the swimmer, as is expected.

The advantage of our raspberry-swimmer description over previously introduced models~\cite{Hernandez-Ortiz05,saintillan07,nash08} is that we obtain the hydrodynamic mobility tensor of our swimmers directly from the LB coupling, which ensures that the raspberry swimmers display the correct translational and rotational behavior in flow. In addition, our raspberries have a finite extent, which leads to a more physical description of particle-particle collisions than can currently be achieved by LB sub-lattice methods~\cite{nash08,nash10}. The coupling to the LB fluid makes it difficult to accurately describe the near-field HI, which is also a limiting factor for other methods. However, our raspberry method ensures that fluid flows around the body, provided that a sufficient number of coupling points is used~\cite{fischer15}. This allows us to capture the higher-order sub-dominant HI terms, which can be of importance for the long-range flow field.

In continuation of this work, we will consider the effect of the shape anisotropy and the quadrupole moment on the motion of self-propelled particles in confining geometries~\cite{degraaf16b}. The presence of such modes will prove crucial to the behavior of such particles, despite their much stronger decay compared to that of the dipole moment. Further extensions of the formalism could include incorporating rotational contributions to the flow field, in a similar spirit the work of Nash~\textit{et al.}~\cite{nash08}, as well as verification of the method for non-axisymmetric shapes. Finally, the characterization method of LF decomposition, as described here, can be applied to determine the HI of complex swimmers, for which only numerical solutions to the flow field exist. We therefore expected that raspberry swimmers and the methods developed in this manuscript will open the way for new directions in the study of active anisotropic particles.

\section*{\label{sec:ack}Acknowledgements}

JdG acknowledges financial support by a NWO Rubicon Grant (\#680501210). JdG and CH thank the DFG for funding through the SPP 1726 ``Microswimmers -- From Single Particle Motion to Collective Behavior''. AJTMM and TNS acknowledge financial support from an ERC Advanced Grant MiCE (291234). TNS thanks EMBO for funding through (ALTF181-2013). We are also grateful to J. Stenhammar, G. Rempfer, and O.A. Hickey for useful discussions.

\end{document}